\documentclass[3p,times,twocolumn]{elsarticle}

\usepackage{ecrc}

\volume{00}

\firstpage{1}

\journalname{}

\runauth{}

\jid{nuphbp}

\jnltitlelogo{Nuclear and Particle Physics Proceedings}

\usepackage{amssymb}
\usepackage[thickspace,amssymb]{SIunits}

\usepackage[figuresright]{rotating}

\begin{document}

\begin{frontmatter}



\dochead{}

\title{Cosmic ray physics with ARGO--YBJ}


\author{Paolo Montini\fnref{fn1} for the ARGO--YBJ collaboration}
%\ead{paolo.montini@roma2.infn.it}
\address{Istituto Nazionale di Fisica Nucleare - Sezione di Roma Tor Vergata, Via della Ricerca Scientifica 1, 00133 Roma (Italy)}
\fntext[fn1]{Now at Dipartimento di Fisica - Sapienza Universit\`a di Roma and INFN sezione di Roma}

\begin{abstract}
The ARGO--YBJ experiment has been in stable data taking for more than five years at the Yangbajing cosmic ray observatory (Tibet, P.R. China, 4300 m a.s.l.). The detector collected about $5\times10^{11}$ events in a wide energy range from few TeVs up to the PeV region. In this work we summarize the latest results in cosmic ray physics particularly focusing on the cosmic ray energy spectrum. The results of the measurement of the allÑparticle and proton plus helium energy spectra in the energy region between  $10^{12} - 10^{16}$ eV are discussed. A precise measurement of the cosmic ray energy spectrum and composition in this energy region allows a better understanding of the origin of the knee and provides a powerful cross--check among different experimental techniques.
\end{abstract}


\end{frontmatter}

% \linenumbers

\section{Introduction}
\label{intro}
The ARGO--YBJ experiment is a full--coverage EAS detector that has been in operation for more than five years at the Yangbajing International Cosmic Ray Observatory (Tibet, P. R. China, 4300 m a.s.l.). The detector has been designed in order to face several open problems in cosmic ray physics and $\gamma$--ray astronomy. The detector operated simultaneously as a wide field of view $\gamma$ ray telescope at the TeV region and  as a high resolution cosmic ray (CR) detector in a wide energy range from TeV up to 10 PeV.  The full--coverage technique, combined with high--altitude operation and high segmentation, allow the detection of showers produced by primaries in the TeV region, so far investigated only by satellite or balloon--borne experiments. \\
In this paper the latest results obtained by ARGO--YBJ  on CR physics are briefly summarized and discussed.

\section{The ARGO--YBJ experiment}
\label{Argo}
The ARGO--YBJ experiment is a full--coverage detector made of a single layer of resistive plate chambers (RPCs) with $\sim 93\%$ active area, surrounded by a partially instrumented ($\sim 20\%$) guard ring. The 1836 RPCs are arranged in 153 clusters each made of 12 chambers. The detector has been equipped with two independent readout systems: each RPC is simultaneously read--out by 80 copper strips $(6.75 \times 61.80\, \centi\meter^2)$ logically arranged in 10 independent pads $(55.6 \times 61.8\, \centi\meter^2)$ and by two large pads called Big Pads  $(139 \times 123\, \centi\meter^2)$. Each strip represents the space granularity of the detector, i.e., the pixel used to sample the particles of the shower front \cite{argo2006}. Each pad signal is sent to a time--to--digital converter and represents the time pixel, allowing a resolution of about $1.8\, \nano \second$ in measuring the particle arrival time.  The installation of the central full--coverage carpet was completed in June 2006. The guard ring was completed during spring 2007 and connected to the DAQ system in November 2007. A trigger logic based on the time coincidence between the pad signals have been implemented. The detector has been in operation from November 2007 up to February 2013 with a trigger threshold $N_{pad} = 20$, corresponding to a trigger rate of about $\sim 3.5\, \kilo\hertz$ with a dead time $\sim 4\%$. % dedicate calibration procedure was implemented in order to achieve high pointing accuracy and angular resolution, which is $\sim 0.4^\circ$ for events with at least 500 fired pads. 
The Big Pads collect the total charge developed by the particles hitting the detector surface and extend the detector operating range up to the PeV region \cite{analog-paper,analog-cina}.  The whole system can be operated at eight different gain scales (G0, $\ldots$, G7) thus extending the detector operating range up to $\sim 10 \, \mathrm{PeV}$. Data from the highest gain scale (G7) have been used for calibration purposes. The intermediate gain scale (G4) overlaps with the digital readout data in a wide energy range between 10 and 100 TeV, providing a cross--calibration of the two techniques. Data from lowest gain scales (G1 and G0) allow the detection of showers with more than $10^4$ particles/$\mathrm{m^2}$ in the core region. 

 \section{Cosmic ray energy spectrum}
 The all--particle energy spectrum of cosmic rays can be roughly described as a single power law with a \emph{knee} at energies around 3.5 PeV. Supernova remnants (SNR) are commonly identified as the sources of Galactic cosmic rays up to the knee region. The origin of the knee and the transition between galactic and extragalactic origin are still under discussion. In the standard picture the origin of the knee is related to a decrease of the flux of protons and He nuclei \cite{KascadeKnee}. Several experiments reported an evidence that the knee of the all--particle spectrum is due to nuclei heavier than Helium. The determination of the elemental composition around the knee therefore plays a key role in the understanding of the origin and acceleration of cosmic rays. The ARGO--YBJ experiment is able to explore the energy region from few TeV up to several PeV.  Measurements of the all--particle and light component (protons plus Helium nuclei) energy spectra are currently under way in this energy range. In order to explore such a wide energy range different approaches have been used:
\begin{enumerate}[-]
\item \emph{Digital Analysis}. It is based on the RPC digital readout data (i.e. the strip multiplicity) and covers the $3-300$ TeV energy range \cite{argoPRD, argoPRD15}.
\item \emph{Analog  Analysis}. It uses the information coming from the RPC analog readout and explores the $30-30000 $ TeV energy range. In this case two approaches have been followed starting from the observed particle distribution at ground level: energy reconstruction on a statistical basis using a bayesian unfolding technique \cite{paoloICRC} and energy reconstruction on an event by event basis \cite{demitriICRC}.
\item \emph{Hybrid Analysis}. It combines the data coming from ARGO--YBJ and a wide field of view Cerenkov telescope \cite{ibrido1,ibrido2}. 
\end{enumerate} 

\subsection{Digital analysis}
\begin{figure}[t]
\includegraphics[width = 0.5\textwidth]{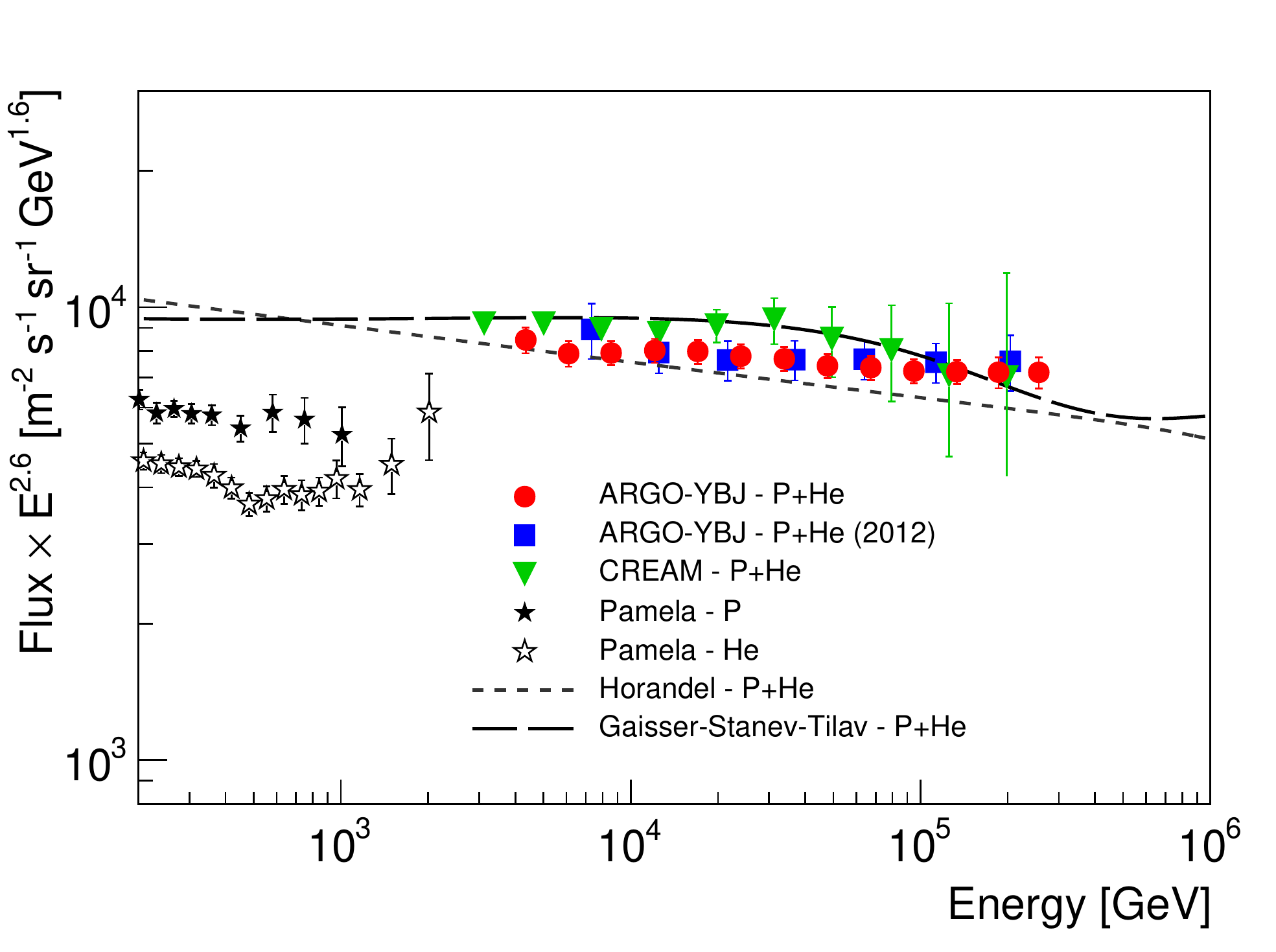}
\caption{The light--component spectrum measured by ARGO--YBJ in the 3--300 TeV energy range compared with other experimental results. The ARGO 2012 data refers to the first results published in \cite{argoPRD}. Results from PAMELA \cite{pamela}, CREAM \cite{cream-apj2011} and the models by H\"orandel \cite{horandel2003} and Gaisser, Stanev and Tilav \cite{GST} are also shown.}
\label{fig:digital}
\end{figure}

The analysis have been performed on  the data collected during the period January 2008 -- December 2012. As widely described in \cite{argoPRD, argoPRD15}, by requiring quasi--vertical $(\vartheta \leqslant 35^\circ)$ showers in an area of about $40 \times 40 \, \mathrm{m^2}$ centered on the detector and applying a selection criteria based on the lateral particle density, a sample of showers mainly produced by light elements has been selected. The energy spectrum has been reconstructed starting from the multiplicity distribution by using an bayesian unfolding technique \cite{dago,thepaper}. The spectrum measured by ARGO--YBJ is shown in figure \ref{fig:digital}. The value of the spectral index of a power--law fit to the ARGO--YBJ data is $-2.64\pm0.01$. The ARGO--YBJ data are in good agreement with the CREAM proton plus helium spectrum. At energies around 10 TeV and 50 TeV the fluxes differ by about 10\% and 20\%, respectively. This analysis demonstrates the excellent stability of the detector over a long period. For the first time a ground--based measurement of the CR spectrum overlap with the results obtained by balloon--borne experiments.

\subsection{Analog analysis}
The RPC charge readout of the ARGO--YBJ experiment allows the measurement of the particle density in the shower core region up to $10^4$ particles/$\mathrm{m}^2$. This system allows the detection of CR in the PeV energy range and therefore the extension of the CR spectrum measurements up to the highest energies.  The high segmentation of the whole system allows a detailed study of the lateral distribution of particles, which can be exploited in order to discriminate among showers produced by primaries of different masses. 
\subsubsection{Unfolding of the CR spectrum}
\begin{figure}[t]
\begin{center}
\includegraphics[width = 0.35\textwidth]{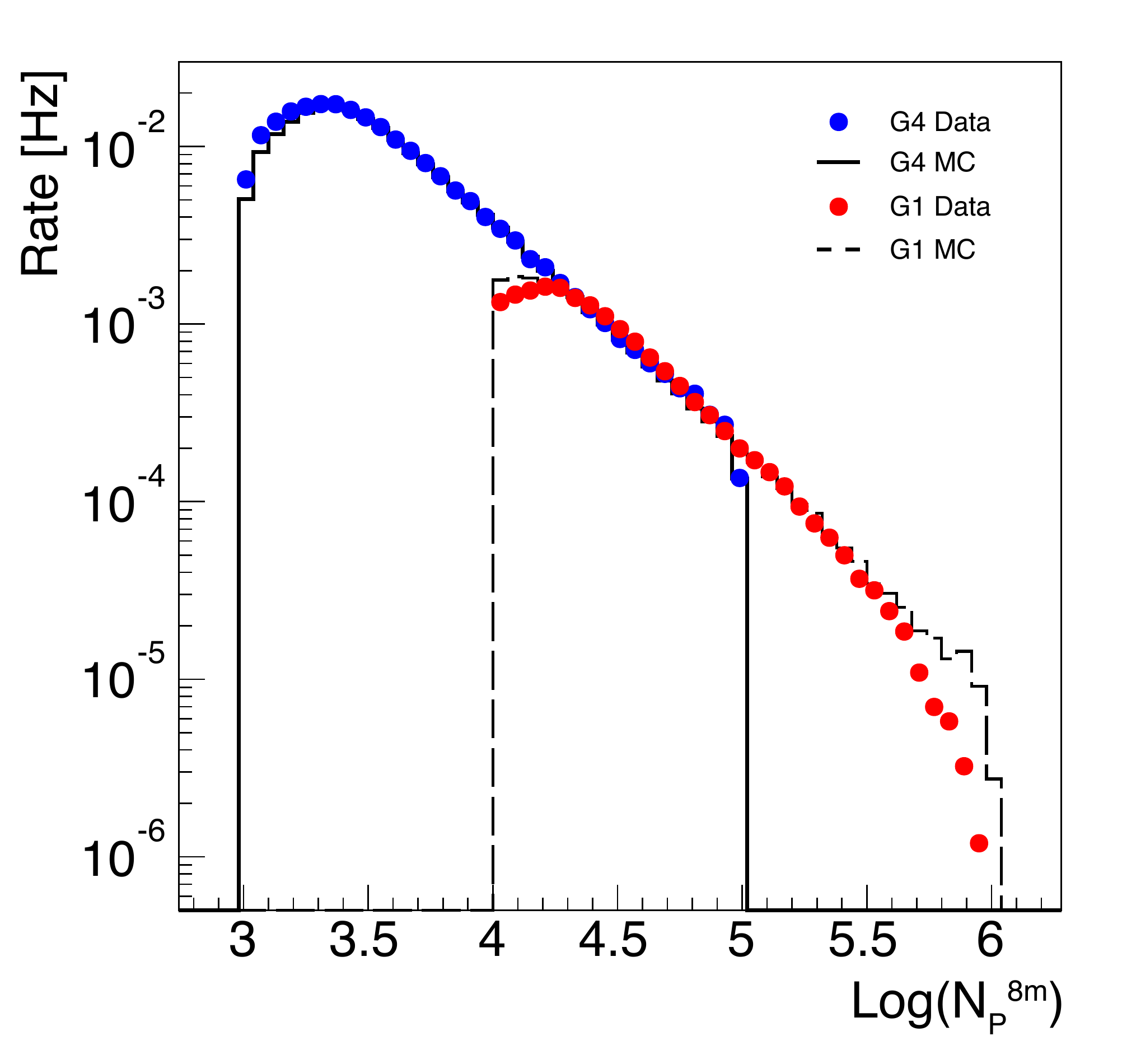}
\caption{Distribution of $N_P^{8m}$ for G4 and G1 datasets.}
\label{fig:np8}
\end{center}
\end{figure}

\begin{figure}[t]
\begin{center}
\includegraphics[width = 0.38\textwidth]{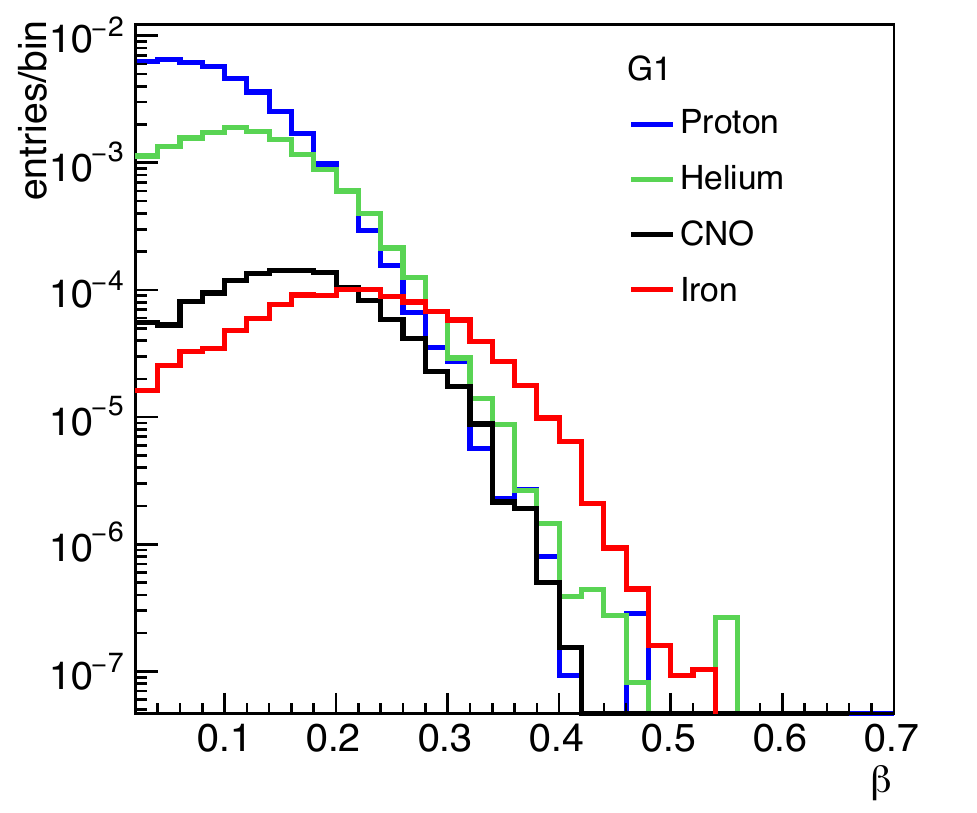}
\caption{Distribution of $\beta$ for G1 dataset.}
\label{fig:beta}
\end{center}
\end{figure}

\begin{figure}[h]
\begin{center}
\includegraphics[width = 0.47\textwidth]{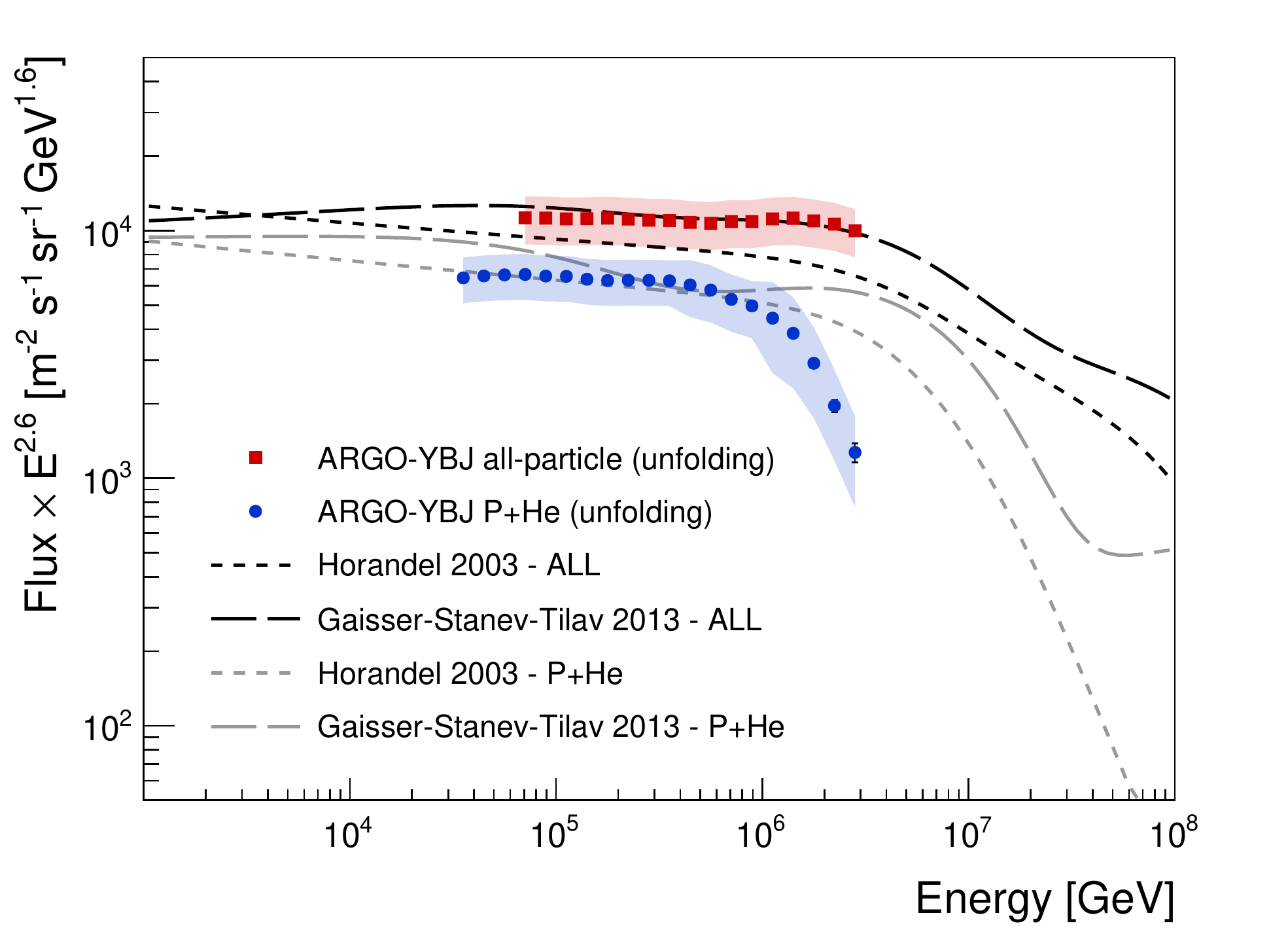}
\caption{All--particle and proton plus helium energy spectra measured by ARGO--YBJ using a bayesian unfolding technique. The systematic uncertainty on the flux is shown by the shaded area and the statistical one by the error bars.}
\label{fig:argobayes}
\end{center}
\end{figure}

As a first step a detailed study of simulated showers have been performed in order to identify an energy estimator and a suitable set of discrimination parameters. The number of particles within 8 meters from the core position $(N_P^{8\mathrm{m}})$ appears to be a robust energy estimator, not affected by bias effects due to the finite detector size. Quasi--vertical $(\vartheta \leqslant 35^\circ)$ showers have been selected inside a fiducial area of about $40\times40\, \mathrm{m^2}$. The analysis has been performed on the data samples collected using the G4 and G1 gain scales, corresponding to an energy range between 10 TeV and 3.5 PeV. In figure \ref{fig:np8} the distribution of $N_P^{8m}$ of the selected events is reported for both G4 and G1 Monte Carlo and experimental data samples. The plot shows a good agreement between experimental data and simulations, therefore demonstrating the reliability of the simulation of the detector response. 
The high segmentation of the detector allows a high--precision study of the lateral particle distribution   at several distances from the core. In showers produced by protons and helium nuclei the highest fraction of particles is localized at small distances from the core, while showers initiated by elements heavier than helium have a considerable fraction of particles even at large distance from the core. The ratio between the particle density measured at several distances from the core and the one measured around the core can therefore be exploited in order to identify light primaries. The quantity $\beta= \rho_5/\rho_0$, where  $\rho_5$ and $\rho_0$ are respectively the particle density measured at $\sim 5\, \mathrm{m}$ from the core and in a region of $\sim 1\, \mathrm{m^2}$ around the core, has been used as discrimination parameter. In figure \ref{fig:beta} the distribution of $\beta$ is reported for different primaries. The plot shows that a large fraction of protons and helium nuclei have small values of $\beta$, demonstrating the possibility of selecting a sample of showers mainly produced by light primaries. 
 The all--particle and light component spectra have been obtained starting from the measured $(N_P^{8\mathrm{m}})$ distribution and the particle densities at different distances from the shower axis by using a bayesian unfolding method. \\
Results are reported in figure \ref{fig:argobayes}.  The all--particle energy spectrum spans the 70--3000 TeV energy range, showing a good agreement with the results of other experiments and therefore demonstrating the reliability of the method. The p+He energy spectrum spans the energy range between 30 TeV and 3 PeV, overlapping the results obtained by analyzing the digital readout data sample. These results are fairly consistent between each other, both concerning the spectral index and flux intensity, demonstrating the reliability of the response of the analog readout system. A deviation from a single power law is clearly evident at energies of about 700 TeV, where a knee--like structure is observed. % The result is affected by a systematic uncertainty of about 15\%. The points at the highest energies are affected also by a contamination of elements heavier than helium not larger than 10\%.  

\subsubsection{Event--by--event energy reconstruction}
\begin{figure}[h]
\begin{center}
\includegraphics[width = 0.4\textwidth]{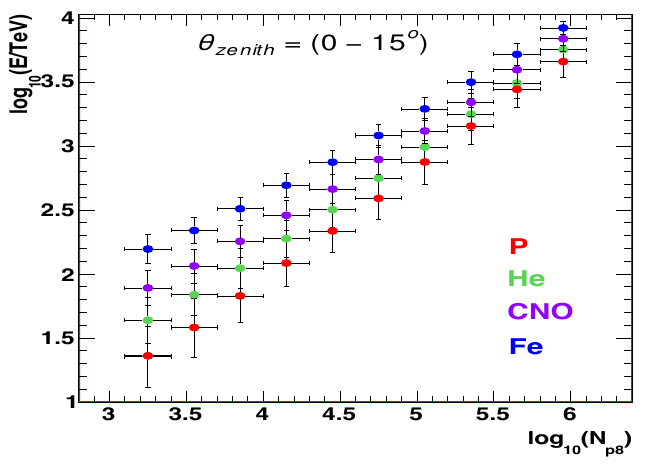}
\caption{Primary energy as a function of  $N_P^{8m}$ for showers produced by different elements.}
\label{fig:Enp8}
\end{center}
\end{figure}

\begin{figure}[h]
\begin{center}
\includegraphics[width = 0.5\textwidth]{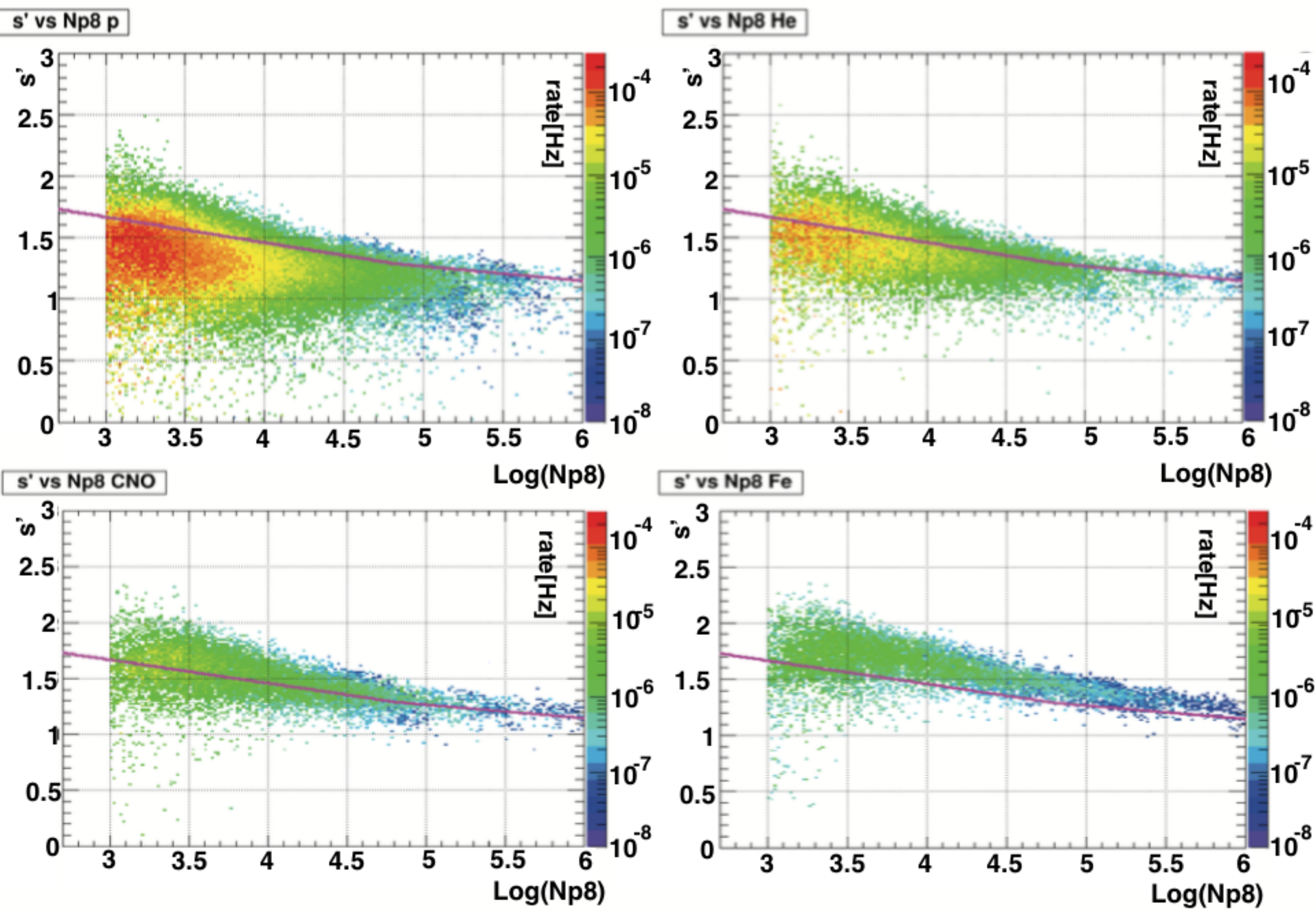}
\caption{The LDF slope $s'$ as a function of the truncated size Np8 as reconstructed for showers initiated by different primary nuclei, as indicated in the upper left labels. The p+He selection cut is shown by the pink lines.}
\label{fig:leccesel}
\end{center}
\end{figure}

\begin{figure}[h]
\begin{center}
\includegraphics[width = 0.45\textwidth]{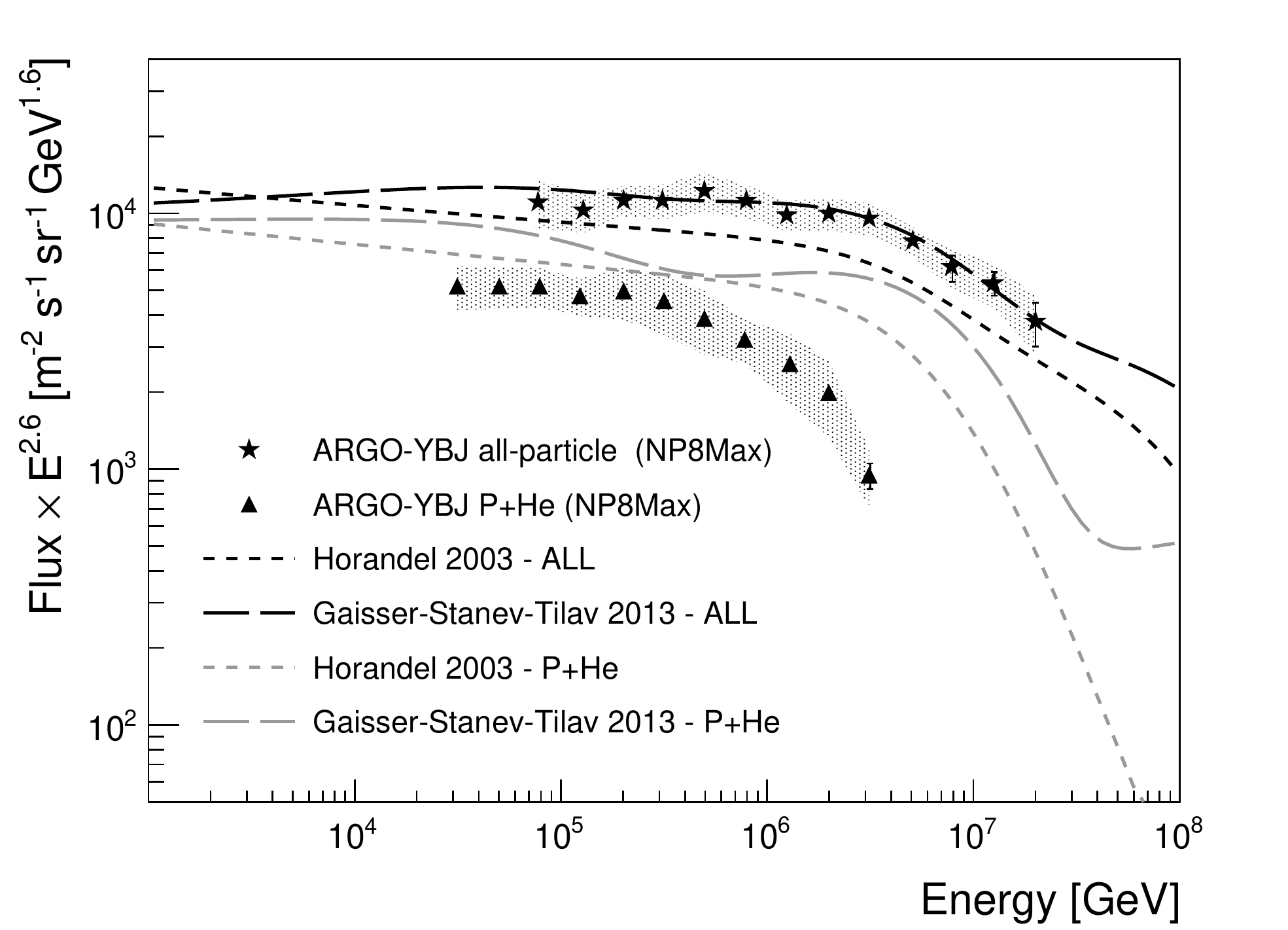}
\caption{All--particle and proton plus helium energy spectra measured by ARGO--YBJ by using an event--by--event energy reconstruction method. The systematic uncertainty on the flux is shown by the shaded area and the statistical one by the error bars.}
\label{leccespec}
\end{center}
\end{figure}

The particle lateral density function close to the shower axis provides information on the shower longitudinal profile in the atmosphere, that can be used in order to estimate the shower age which is related to $X_{max}$, the atmospheric depth at which the shower reaches it maximum size. The combined use of the shower energy and age estimation gives a sensitivity to the primary mass and therefore the possibility of selecting a p+He sample. As shown in figure \ref{fig:Enp8} the truncated size is a mass--dependent energy estimation parameter.
In order to have a mass--independent energy estimator the LDFs of individual showers have been fitted up to 10 m from the core with a suitable function to get the slope $s'$.  The LDF slope $s'$, combined with the measurement of $N_P^{8m}$ can give information primary particle nature. From these studies we find that, for a given primary, the $s'$ value decreases when $N_P^{8m}$ (i.e. the energy) increases, this being due to the observation of younger (deeper) showers at larger energies. Moreover, for a given range of $N_P^{8m}$, $s'$ increases going from proton to iron, as a consequence of a larger primary interaction cross-section. This outcome has two important implications: the measurement of $N_P^{8m}$ and $s'$ can both constrain the shower age and give information about the primary mass. By assuming an exponential absorption after the shower maximum we can get the size at maximum $N_{P MAX}^{8m} \approx N_{P}^{8m} \cdot \mathrm{exp}[  (h_0 \sec \vartheta- X_{max}(s'))/\lambda_{abs}]$. A suitable choice of the absorption lenght allows to get a mass--independent value of $N_{P MAX}^{8m}$ which is a suitable energy estimator \cite{bernaICRC}. Selecting quasi-vertical events $(\vartheta \leqslant 15^\circ)$ with different values of the truncated size $N_P^{8m}$ with the described procedure we reconstructed the CR all-particle energy spectrum shown in the Fig.4 in the energy range 80 TeV--20 PeV.  The overall systematic uncertainty is shown by the shaded area while statistical uncertainty is shown by error bars. Starting from the data sample selected for the all--particle spectrum analysis  a selection has been made in order to have a sample of p and He initiated showers. In figure \ref{fig:leccesel} the values of $s'$ are reported as a function of $N_P^{8m}$ as reconstructed from simulated showers initiated by different primaries.  The line in the plots shows the cut used in selecting the p+He enriched sample from real data. The efficiency in selecting p and He initiated showers and the  heavier elements contamination are at the level of 90\% and 10\% respectively, with variations of few percent depending the energy region and the adopted flux parametrizations.
Taking into account these values (and their energy dependence), the p+He flux has been obtained. The result is shown in figure \ref{leccespec}. The systematic uncertainty on the flux is shown by the shaded area and the statistical one by the error bars. 

\subsection{Hybrid analysis}
\begin{figure}[h]
\begin{center}
\includegraphics[width = 0.38\textwidth]{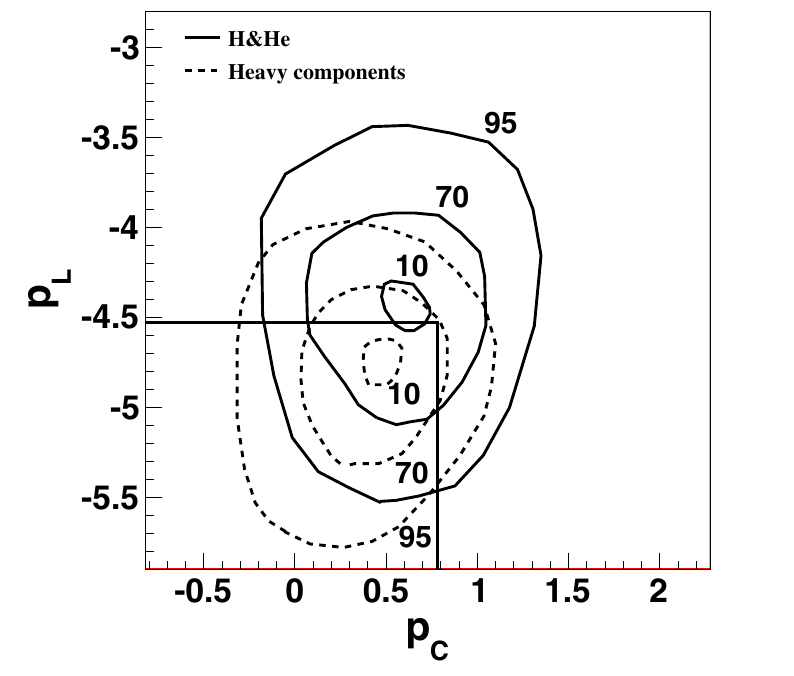}
\caption{Composition-sensitive parameters $p_L$ and $p_C$ for two mass groups, p+He (solid contours) and heavier masses (dashed contours) including 1:1:1 mixing of CNO, MgAlSi, and iron. The primary energy of the plotted events is between 100 TeV and 10 PeV. Numbers on the contours indicate the percentage of contained events.}
\label{fig:hybridsep}
\end{center}
\end{figure}

\begin{figure}[h!]
\includegraphics[width = 0.45\textwidth]{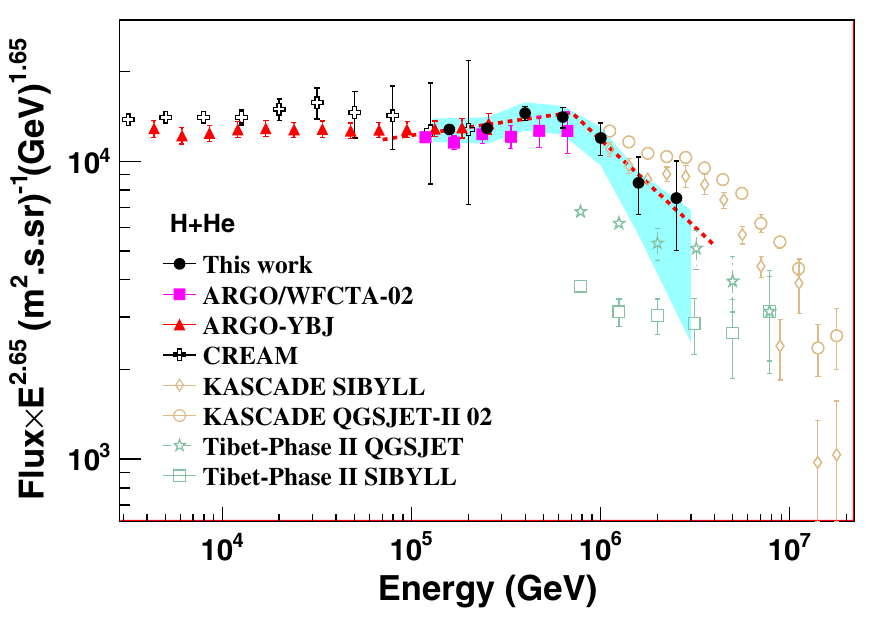}
\caption{ H\&He spectrum obtained by the hybrid experiment with ARGO-YBJ and the imaging Cherenkov telescope.
A clear knee structure is observed around 700 TeV. The p+He spectra by CREAM \cite{cream-apj2011}, ARGO-YBJ \cite{argoPRD15} and the hybrid experiment \cite{CPC} below the knee, the spectra by Tibet AS \cite{amenomori2011cosmic} and KASCADE \cite{apel2013kascade} above the knee are shown for comparison. In our result, the error bar is the statistical error, and the shaded area represents the systematic uncertainty}
\label{hybridspect}
\end{figure}

\begin{figure}[h!]
\begin{center}
\includegraphics[width = 0.5\textwidth]{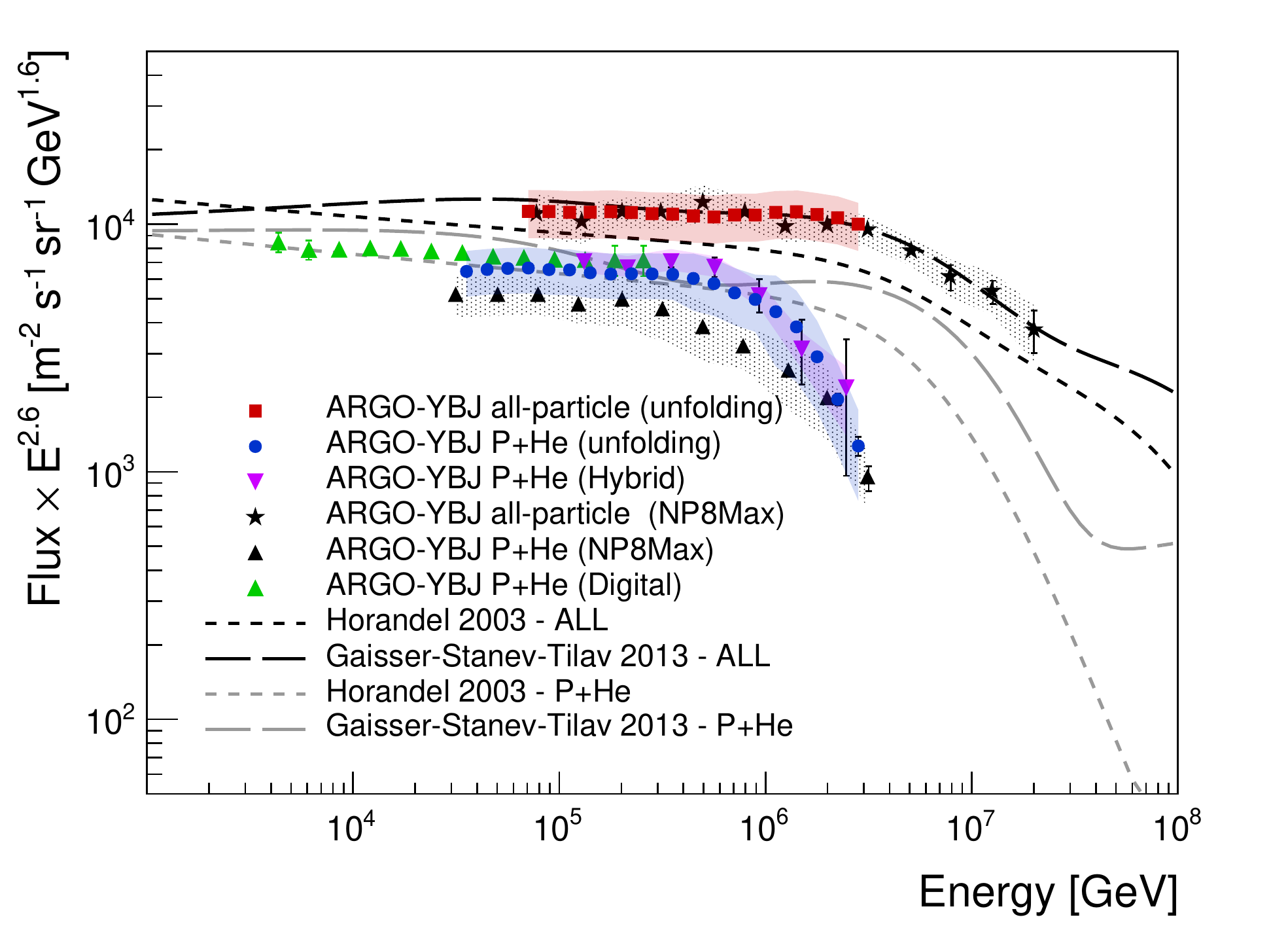}
\caption{ The all-particle and proton plus helium energy spectra measured by ARGO--YBJ by using different experimental techinques.}
\label{fig:argoAll}
\end{center}
\end{figure}

The measurement of the p+He spectrum has been carried out by using an hybrid technique which combines the information coming from the ARGO--YBJ detector and a prototype of the wide FoV Cherenkov telescope array (WFCTA) of the LHAASO project \cite{zhen2010future}. The telescope is located at about 79 m far from the ARGO--YBJ detector center. The Cherenkov telescope consists of an array of $16\times16$ PMTs and has a field of view of $14^\circ \times  16^\circ$ with a pixel size of $\sim 1^\circ \times  1^\circ$.  The idea is to combine two mass sensitive parameters: the particle density near the shower core measured by the analog readout of ARGO--YBJ and the shape of the Cherenkov image measured by WFCT.  A total exposure time of $7.28 \times 10^5$ seconds has been obtained between December 2010 and February 2012.  A sample of about 8700 events above 100 TeV have been selected according to the selection criteria described in \cite{ibrido2}.
%A sample of about 8700 events above 100 TeV have been selected according to the following selection criteria: reconstructed core position inside the full coverage carpet of ARGO--YBJ (excluding an outer region 1 m large), more than 1000 fired pads in ARGO--YBJ, a space angle between the shower direction and the telescope main axis less than $6^\circ$, more than six fired pixels in the WFCT PMT matrix. These criteria ensures an angular resolution of $0.3^\circ$ and a core position resolution of about 2 m. 
According to MC simulations the largest number of particles detected by the RPC carpet $(N_{max})$ gives an estimate of the particle density in the core region (i.e. within 3 m). For a given energy in showers produced by light nuclei $(N_{max})$ is expected to be larger than in showers produced by heavy particles and can be used in order to select different primary masses. In addition $N_{max} \propto N_{pe}^{1.44}$, where $N_{pe}$ is the total number of photoelectrons collected by the Cherenkov telescope. A new parameter $p_L = \log_{10}(N_{max})-1.44 \cdot \log_{10}(N_{pe})$ can be defined in order to describe the correlation between $N_{max}$ and $N_{pe}$. The Cherenkov image of a shower can be described by using the Hillas parameters \cite{hillas}. Showers which develop higher in atmosphere, like iron--induced events, have a more stretched Cerenkov image (i.e. narrower an longer)  with respect to young showers produced by light particles. The ratio between length and width $(L/W)$ is a good estimator of the primary composition. The ratio $L/W$  is also proportional to the shower impact parameter $R_p$ which is the distance between the telescope and the shower core position. The variable $p_C = L/W - R_p/109.9\, \mathrm{m} -0.1\cdot \log_{10 }(N_{pe})$ has been introduced. The combination of $p_L$ and $p_C$ allows the selection of a sample of p and He induced showers. In figure \ref{fig:hybridsep} a contour plot of the $p_L-p_C$ map is reported for different primaries in the energy range 100 TeV -- 10 PeV. The plot shows the possibility of selecting a sample of p and He induced showers with high efficiency. 
The energy resolution is about 25\%, nearly constant throughout the energy range from 100 TeV up to 3 PeV, with a systematic uncertainty on the energy scale of 9.7\%. The resulting p+He energy spectrum is reported in figure \ref{hybridspect}. A knee--like feature is observed at energies around 700 TeV, with a statistical significance of 4.2 standard deviations. \\
All the ARGO--YBJ results concerning the all--particle and p+He spectra are summarized in figure \ref{fig:argoAll}.

 \section{Cosmic ray anisotropy}
\begin{figure}
\includegraphics[width = 0.48\textwidth]{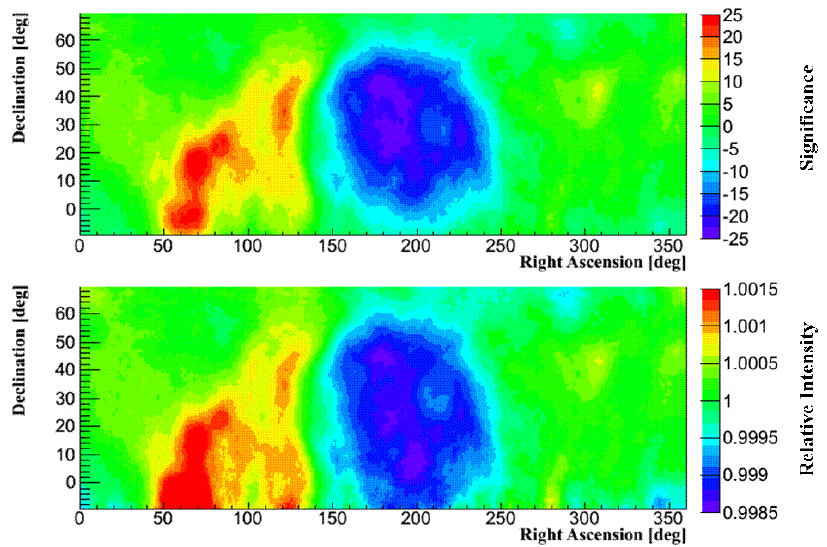}
\caption{Significance(upper panel) and relative intensity (lower panel) maps measured by ARGO--YBJ in the equatorial coordinate system.}
\label{fig:aniso}
\end{figure}

\begin{figure}[h]
\includegraphics[width = 0.48\textwidth]{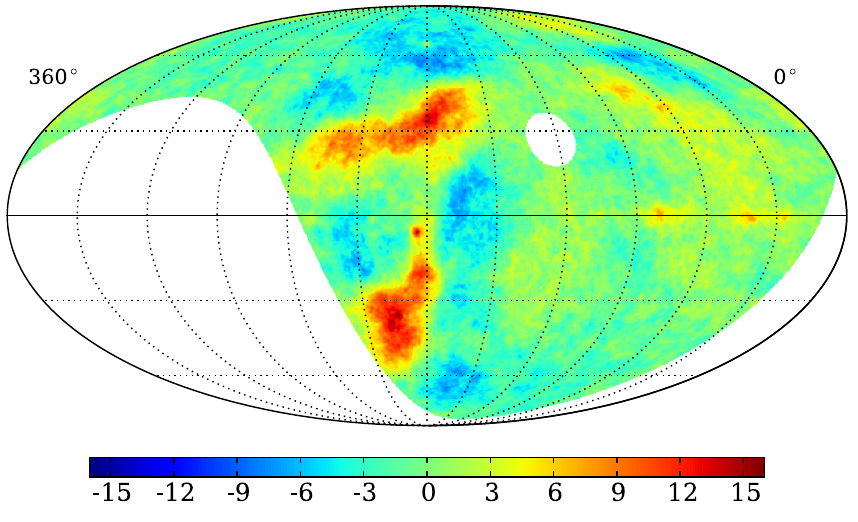}
\caption{ARGO--YBJ sky map in galactic coordinates. The color scale gives the statistical significance of the observation. The map center points towards the galactic anticenter.}
\label{fig:mediumaniso}
\end{figure}

The arrival directions of cosmic rays are highly randomized by the interaction with the galactic magnetic field. The observed flux is therefore expected to be isotropic. Several experiments, however, observed an energy dependent \emph{large scale} anisotropy (LSA) in the sidereal time frame which suggests the existence of two different broad regions. The first region (\emph{tail--in}) shows an excess of CRs and distributed around $40^\circ$ and $90^\circ$ in Right Ascension (R.A.), while the second (\emph{loss--cone}) shows a deficit distributed around $150^\circ$ and $240^\circ$ in R.A. The amplitude is about $10^{-4} - 10^{-3}$. The ARGO--YBJ measurement of the LSA is reported in figure \ref{fig:aniso}. The \emph{loss--cone} and \emph{tail--in} regions are clearly visible with a significance greater than 20 s.d. 
The CR anisotropy has been studied in seven energy bins and the corresponding R.A. profiles have been fitted with the first two harmonics. The amplitude and phase of the first harmonic have been studied as a function of energy. The ARGO--YBJ results are in good agreement with those obtained by other experiments, suggesting a decrease of the amplitude of the first harmonic at energies greater than $10 \,\tera \electronvolt$ (see \cite{ArgoLargeAniso} for details). In the last years many experiments presented an evidence of the existence of a medium angular scale anisotropy in both emispheres. In figure \ref{fig:mediumaniso} the ARGO--YBJ sky map of medium angular scale ($\sim 10^\circ$) in equatorial coordinates is reported \cite{ArgoMediumAniso}. The median energy of the proton flux has been estimated from Monte Carlo simulations and turns out to be $E_P^{50} = 1.8\, \tera\electronvolt$.  No gamma/hadron separation has been applied therefore the map is filled with all CR including photons. The analysis have been performed on about 4.5 years of data. The most evident structures are localized around the positions $\delta \sim 40^\circ,\,  \alpha \sim 120^\circ$ and  $\delta \sim -5^\circ,\,  \alpha \sim 60^\circ$. These regions have been observed with a statistical significance of $\sim 15$ s. d. and are consistent with the regions detected by the Milagro experiment \cite{MilagroAniso}. The right side of the map shows several few degree excesses that cannot be addressed as random fluctuations (the significance is larger than 7 s. d.). As discussed in \cite{ArgoMediumAniso} these structures have been reported for the first time by ARGO--YBJ.

\subsection{Anisotropy of light elements}
Experimental results from EAS--TOP \cite{eastop} and ICECUBE \cite{icecube-aniso} demonstrate that the morphology of the cosmic ray anisotropy changes at energies around 100 TeV.  At energies below 100 TeV the anisotropy is dominated by dipole and quadrupole components. The non--dipolar structure of the anisotropy at higher energies challenges the current paradigm of CR diffusion models. The ICETOP experiment has shown that the structure observed at $\sim$ 400 TeV persists also at PeV energies \cite{aniso-icetop} with a deeper deficit. The origin of the strenght of this deficit is still unclear. It can be related to propagation effects from a given source or to the contribution of heavier nuclei at energies around the knee. As a consequence the measurement of the contribution of individual elements to the total CR anisotropy around the knee would clearly provide fundamental information in the understanding of the propagation of CR in our galaxy.\\
The ARGO--YBJ collaboration has recently presented a preliminary analysis of the Galactic CR anisotropy for events induced by protons and Helium nuclei \cite{lightAniso}. \\
%In anisotropy studies the crucial point is to estimate the number of background events (i.e. the number of events expected in case of complete isotropy). This quantity is related to the total CR flux, to the effective area and to exposure time. Since the CR flux can be assumed as constant in time all the variations of the background can be addressed to variation of effective area that 
In order to evaluate the contribution due to light nuclei to the total CR anisotropy a new method has been presented. The basic idea of this new approach is to measure the anisotropy of a particular subset of events (in this case protons and helium nuclei) taking the complete set of events as an estimator of the background. Several consistency checks have been made on a sample of simulated data. %The analysis has been performed on $\sim 5\times10^{11}$ events collected in the period between January 2008 and December 2012. Showers produced by light elements have been selected according to the same criteria used in the digital analysis. In figure \ref{fig} the sky map obtained in this analysis is reported. Although statistic is really poor the fit with a dipole function of the R.A. profile gives a phase at $-6^\circ \pm 10^\circ$, in coincidence with the loss cone. 
This analysis showed an interesting potential of searching possible correlations between the known large scale anisotropy structures and deviations from isotropy induced by light elements.

 \section{Conclusions}
 The ARGO--YBJ detector has been in stable data taking in its full configuration for more than five years at the Yangbajing International Cosmic Ray Observatory. With a duty cycle of $\sim 86\%$ the detector has collected more than $5\times 10^{11}$ events. The peculiar characteristics of the detector like high segmentation and high altitude operation allows the detection of showers in a wide energy range between 1 TeV and 10 PeV. The analog readout system provides a powerful tool to study the particle distribution at several distances from the core. The detector has demonstrated an excellent stability over a long period. Some important achievements  on cosmic ray physics so far obtained by the ARGO--YBJ experiment have been summarized in this paper. The cosmic ray energy spectrum has been investigated in a wide energy range using three different approaches. The all--particle spectrum is consistent with previous experimental observations. The energy spectrum of light elements (protons plus helium nuclei) has been measured from 3 TeV up 3 PeV.  A gradual change of the spectral index at energies around 700 TeV is clearly evident.  This result demonstrates the possibility of exploring the cosmic ray properties in a wide energy range with a single groundÐ based experiment and opens new scenarios about the evolution of the light component energy spectrum towards the highest energies and the origin of the knee. \\
Both the large and medium scale have been investigated by the ARGO--YBJ experiment. The measurement of the anisotropy for different primary particle masses in the knee energy region should be a high priority of the next generation ground-based experiments to discriminate between different propagation models of CRs in the Galaxy. A preliminary analysis of the CR anisotropy for (p+He)-induced events has been carried out with the ARGO-YBJ experiment developing a new analysis method.


%% References

%\nocite{*}
\bibliographystyle{elsarticle-num}
\bibliography{ARGO-Analog}






% \end{thebibliography}

\end{document}